\documentclass[twocolumn,showpacs,prl]{revtex4}
\usepackage{amssymb}
\usepackage{makeidx}
\usepackage{amsmath}
\usepackage{graphicx}
\usepackage{dcolumn}
\usepackage{bm}

\begin{document}

\title{Quasimonoenergetic and low emittance ion bunch generation from
ultrathin targets by counterpropagating laser pulses of ultrarelativistic
intensities}
\author{H.K. Avetissian}
\email{avetissian@ysu.am}
\author{A.K. Avetissian}
\author{G.F. Mkrtchian}
\author{Kh.V. Sedrakian}
\affiliation{Centre of Strong Fields Physics, Yerevan State University, 1 A. Manukian,
Yerevan 0025, Armenia }
\date{\today }

\begin{abstract}
A new method for generation of quasimonoenergetic and low emittance fast
ion/nuclei bunches of solid densities from nanotargets by two
counterpropagating laser pulses of ultrarelativistic intensities is
proposed, based on the threshold phenomenon of particles "reflection" due to
induced nonlinear Compton scattering. Particularly, a setup is considered
which provides generation of ion bunches with parameters that are required
in hadron therapy.
\end{abstract}

\pacs{41.75.Jv, 52.38.Kd, 37.10.Vz, 52.65.Rr}
\maketitle



Acceleration of ions with superstrong laser beams has attracted broad
interest over the last years conditioned by a number of important
applications, such as inertial confinement fusion \cite{fusion}, isotope
production \cite{isotop}, hadron therapy \cite{therap}, and etc. Most of
these mechanisms are based on the indirect processes, when ions are
accelerated by the space-charge fields induced in the laser-target
interaction process. There are several regimes of ions acceleration
depending on the laser intensity: Target Normal Sheath Acceleration regime 
\cite{TNSA}, Coulomb Explosion regime \cite{CE}, Radiation Pressure Dominant
regime \cite{RPDA}, and the shock wave acceleration mechanism \cite{shock}.
Typically in Target Normal Sheath Acceleration regime, the accelerating
gradient reaches of $10^{12}$ $\mathrm{V/m,}$ but due to short interaction
range $\sim $ $1\ \mathrm{\mu m}$ it accelerates the ions to tens of \textrm{%
MeV} energies. The energy distribution of accelerated ions obtained by the
laser plasma interaction is usually exponential with almost 100\% energy
spread up to a cutoff energy. Only within the last years became possible to
produce monoenergetic ion beams from targets using special technique \cite%
{exper}. In these experiments up to $60\ \mathrm{MeV}$ protons was observed.
In Radiation Pressure Dominant regime, with circularly polarized laser
pulses when heating of electrons is reduced, it is expected to achieve 
\textrm{GeV/nucleon }energy\textrm{\ }range at the ultrarelativistic laser
intensities \cite{Circ}.

The use of mentioned indirect processes for ions acceleration, because of
the lack of directly accelerating laser fields of required intensities, has
some undesirable factors, such as uncontrollable beam parameters, etc. The
state-of-the-art laser systems are capable of generating electromagnetic
pulses with intensities exceeding on several orders the threshold of
relativism for electrons. For heavy particles, laser intensities, at which,
e.g., an ion with atomic mass number $\mathcal{A}$ and charge number $%
\mathcal{Z}$ becomes relativistic, is defined by the condition $\Xi \gtrsim
1 $, where $\Xi =\mathcal{Z}eE\lambdabar /\mathcal{A}m_{u}c^{2}$ is
relativistic dimensionless parameter of a wave-particle interaction ($e$ is
the elementary charge, $c$ is the light speed in vacuum, and $m_{u}$ is the
atomic mass unit $m_{u}\simeq 1.66\times 10^{-24}\mathrm{g}$ ) and
represents the work of the field with electric strength $E$ on a wavelength $%
\lambda $ ($\lambdabar =\lambda /2\pi $) in the units of particle rest
energy. Laser intensities, at which relativistic effects become important
for an ion ($\Xi =1$ -threshold value) can be estimated as: $I_{r}=\mathcal{A%
}^{2}\mathcal{Z}^{-2}\times 4.55\times 10^{24}\ \mathrm{W\ cm}^{-2}[\lambda /%
\mathrm{\mu m}]^{-2}$. The availability of such intensities now is in the
scope of ELI project \cite{ELI} and it is of interest the mechanisms for
direct laser acceleration of heavy particles. The spectrum of direct
acceleration mechanisms of nuclei/ions by a single laser pulse is very
restricted, since one should use the laser beams focused to subwavelength
waist radii, or use subcycle laser pulses. Direct acceleration of ions by
petawatt laser beams focused to subwavelength waist radii has been explored
with radially polarized lasers \cite{Salamin}, where ions energies cover the
domain of application in hadron therapy \cite{therap}. However, the
realization of this scheme with radially polarized laser pulses of required
ultrarelativistic intensities is problematic. Hence, it is reasonable (for
low energy spreads and emittances of accelerated ions bunches) to consider
coherent processes of laser-particle interaction. Among those induced
Compton process is practically more effective. Thus, in this process a
critical value of intensity exists \cite{effect} above which the nonlinear
threshold phenomenon of particles "reflection" occurs from the formed moving
wave-barrier which opens a principally new way for particles acceleration on
the short distances -even shorter than a wavelength \cite{Book}. Note,
however, that this phenomenon of critical field, in principle, takes place
for plane waves. How the critical-field-effect will change the interaction
dynamics for strongly nonplane -tightly focused ultrarelativistic laser
pulses, this is a problem for investigation in an each specific case, for
concrete pulse-configurations (see, e.g. \cite{accel}).

In this letter we propose a new method for generation of solid density,
monoenergetic and low emittance ions/nuclei bunches of ultrashort durations.
Accordingly, we will study laser acceleration of ions/nuclei in the induced
Compton process with strongly nonplane ultrashort pulses, taking into
account the specific feature of the critical field for generation of fast
ion beams with low energy spreads and emittances. For realization of the
latter we propose to focus two counterpropagating laser beams of different
frequencies and ultrarelativistic intensities onto an nanoscale-solid-plasma
targets with $\mathcal{Z}/\mathcal{A}=1;1/2$ (for protons and nuclei,
respectively). In particular, we consider the specific range of energies
that is of great interest to hadron therapy \cite{therap} with
quasimonoenergetic-monochromatic ion beams.

At first we consider the classical dynamics of an ion in vacuum at the
interaction with two counterpropagating plane waves of carrier frequencies $%
\omega _{1},$ $\omega _{2}$ (let $\omega _{1}>$ $\omega _{2}$), wavenumbers $%
\mathbf{k}_{1}=\left\{ \omega _{1}/c,0,0\right\} $, $\mathbf{k}_{2}=\left\{
-\omega _{2}/c,0,0\right\} $, and slowly varying electric field amplitudes $%
E_{1}\left( \tau _{1}\right) $, $E_{2}\left( \tau _{2}\right) $ ($\tau
_{1}=t-x/c$, $\tau _{2}=t+x/c$). Both waves are assumed to be linearly
polarized along the $OY$ direction:%
\begin{equation}
\mathbf{E}_{1,2}\left( x,t\right) =\left\{ 0,E_{1,2}\left( \tau
_{1,2}\right) \cos \omega _{1,2}\tau _{1,2},0\right\} .  \label{E12}
\end{equation}%
Dynamics of an ion in the resulting electric $\mathbf{E=E}_{1}+\mathbf{E}%
_{2} $ and magnetic $\mathbf{H=}\widehat{\mathbf{x}}\times \left( \mathbf{E}%
_{1}-\mathbf{E}_{2}\right) $ fields is governed by the equations:

\begin{equation}
\frac{d\mathbf{\Pi }}{dt}=\frac{e\mathcal{Z}}{\mathcal{A}m_{u}c}\left( 
\mathbf{E}+\frac{\mathbf{\Pi }\times \mathbf{H}}{\gamma }\right) ,\ \frac{%
d\gamma }{dt}=\frac{e\mathcal{Z}}{\mathcal{A}m_{u}c}\frac{\mathbf{\Pi }\cdot 
\mathbf{E}}{\gamma }.  \label{eeq}
\end{equation}%
Here we have introduced normalized momentum $\mathbf{\Pi }\ =\mathbf{p/(}%
\mathcal{A}m_{u}c)\ $and energy $\gamma =\sqrt{1+\mathbf{\Pi }^{2}}$
(Lorentz factor). When the particle initial transversal momentum is zero,
and the waves are turned on/off adiabatically at $t\rightarrow \mp \infty $,
from Eq. (\ref{eeq}) for transversal momentum one can obtain (coordinate $z$
is cyclic, so $\Pi _{z}=\mathrm{const}$ $\equiv 0$): 
\begin{equation}
\Pi _{y}\left( \Xi _{1},\Xi _{2}\right) =\Xi _{1}\left( \tau _{1}\right)
\sin \omega _{1}\tau _{1}+\Xi _{2}\left( \tau _{2}\right) \sin \omega
_{2}\tau _{2}.  \label{py}
\end{equation}%
With the help of Eq. (\ref{py}), one can obtain equations for longitudinal
momentum and energy, which include four nonlinear interaction terms: two of
them are proportional to $\Xi _{1,2}^{2}\left( \tau _{1,2}\right) \sin
2\omega _{1,2}\tau _{1,2}$ and describe interaction with separate waves that
can not provide real energy change for ion. The term proportional to $\Xi
_{1}\left( \tau _{1}\right) \Xi _{2}\left( \tau _{2}\right) \sin \left(
\omega _{1}\tau _{1}+\omega _{2}\tau _{2}\right) $ describes interaction
with the fast interference wave making no contribution to real energy
exchange too. This term is responsible for particle-antiparticle pair
production from Dirac vacuum \cite{pair}. The resonant interaction of the
ion for acceleration is governed by the slowed interference wave $\Xi
_{1}\left( \tau _{1}\right) \Xi _{2}\left( \tau _{2}\right) \sin \widetilde{%
\omega }\left( t-x/c\beta _{ph}\right) $, where $\widetilde{\omega }=\omega
_{1}-\omega _{2}$ and $\beta _{ph}=\widetilde{\omega }/\left( \omega
_{1}+\omega _{2}\right) <1$ are the frequency and phase velocity of the
slowed wave. Hence, keeping only this resonant term, one can obtain the
following integral of motion in average:%
\begin{equation}
\gamma -\beta _{ph}\Pi _{x}\simeq \gamma _{0}-\beta _{ph}\Pi _{0x}.
\label{int}
\end{equation}%
From Eqs. (\ref{int}), (\ref{py}) and dispersion relation $\gamma =\sqrt{1+%
\mathbf{\Pi }^{2}}$ one can see that for the certain values of $\Xi _{1,2}$
(which we call critical) the following relation for an ion average
transversal momentum in the field may be satisfied:%
\begin{equation*}
\sqrt{\overline{\Pi _{y}^{2}\left( \Xi _{1},\Xi _{2}\right) }}>\gamma
_{0}\left\vert \beta _{ph}-\beta _{0x}\right\vert \left( 1-\beta
_{ph}^{2}\right) ^{-1/2},
\end{equation*}%
at which the slowed interference wave becomes a potential barrier causing
ion reflection from this moving wave-barrier. After the interaction ($\Xi
_{1,2}=0$) for the reflected ion final energy we have:%
\begin{equation}
\gamma _{f}\simeq \gamma _{0}+2\gamma _{0}\beta _{ph}\left( \beta
_{ph}-\beta _{0x}\right) \left( 1-\beta _{ph}^{2}\right) ^{-1}.  \label{gain}
\end{equation}%
Formula (\ref{gain}) shows that ion acceleration depends neither on the
fields' magnitudes nor the interaction length. Namely this feature of
reflection phenomenon is used here for generation of monoenergetic and low
emittance ions/nuclei bunches of ultrashort durations.

Then for actual strongly nonplane and supershort ultrarelativistic laser
pulses of certain configurations the problem is solved with the help of
particle-in-cell (PIC) simulations. Here we report on the results of the
2D3V PIC simulations of counterpropagating waves interaction with
nanolayers. We use the code XOOPIC, which is a relativistic code based on
PIC method \cite{xoopic}. Then we consider two type of ion targets: ionized
hydrogen $^{1}\mathrm{H}^{+}$ and fully ionized carbon $^{12}\mathrm{C}^{6+}$%
. The simulation box size is $40\lambda _{1}\times 20\lambda _{1}$ in $xy$
plane. Number of cells are $4000\times 200$. The total number of
macro-particles is about $4\times 10^{4}$ for hydrogen target and $2.4\times
10^{5}$ -for carbon target. Target is assumed to be fully ionized. This is
justified since the intensity for fully ionization of carbon is about $%
10^{19}$ $\mathrm{W/cm}^{2}$, while we use in simulations intensities at
least on five orders of magnitude larger. So, the target will become fully
ionized well before the arrival of the pulses' peaks. Electrons and ions are
assumed to be cold in the target, $T_{e}=T_{i}=0$.

\begin{figure}[tbp]
\includegraphics[width=.43\textwidth]{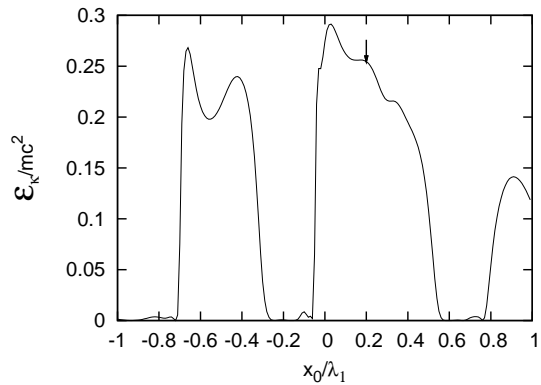}
\caption{The final scaled kinetic energy versus the initial position of the
ion $x_{0}$ in units of wavelength $\protect\lambda _{1}$. The arrow shows
position of the target layer.}
\end{figure}

\begin{figure}[tbp]
\includegraphics[width=.4\textwidth]{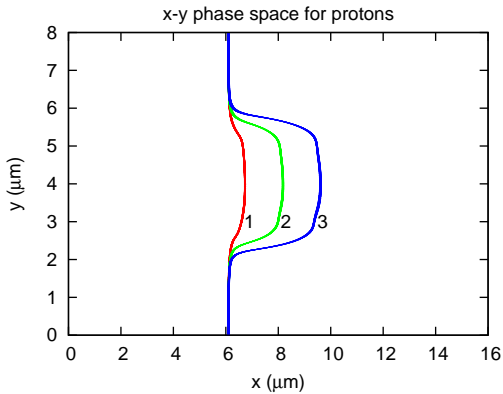}
\caption{(Color online) Ion layer blown out from a thin foil by
counter-propagating waves. The ($x-y$) phase space for protons at instants
(1) $15T_{2}$, (2) $18T_{2}$ and (3) $21T_{2}$.}
\end{figure}

\begin{figure}[tbp]
\includegraphics[width=.43\textwidth]{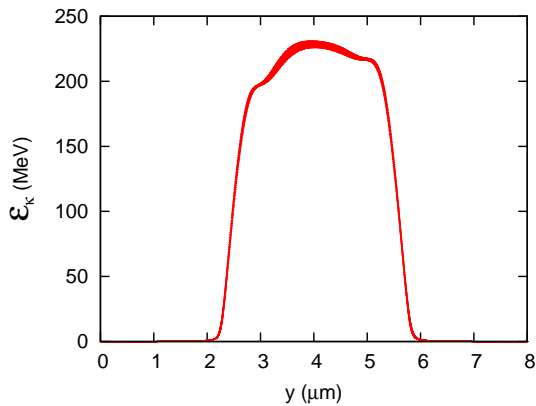}
\caption{(Color online) The final kinetic energy distribution (at instant $%
21T_{2}$) of accelerated protons versus the transversal position $y$.}
\end{figure}

\begin{figure}[tbp]
\includegraphics[width=.43\textwidth]{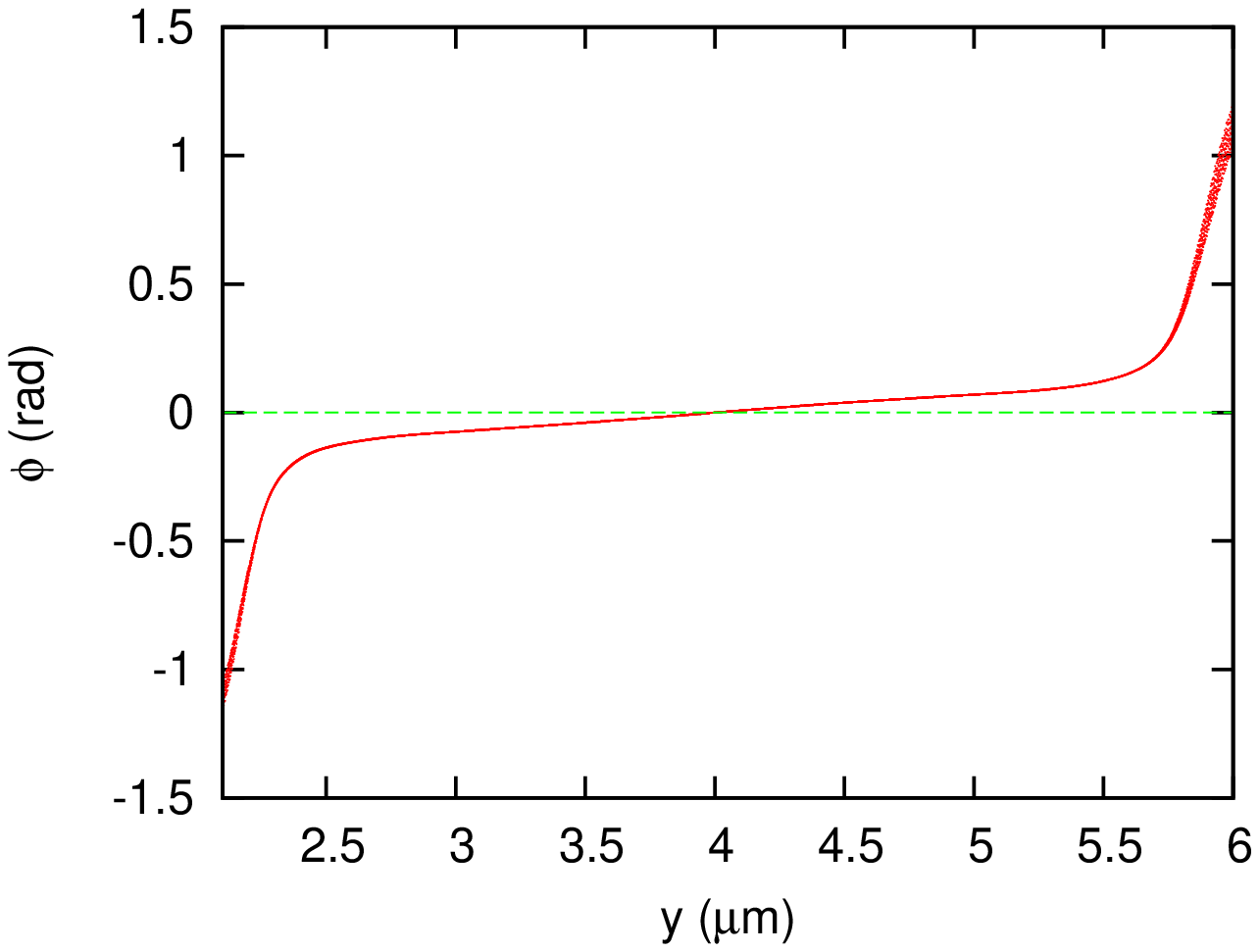}
\caption{(Color online) The angular distribution of accelerated protons
versus the transversal position $y$.}
\end{figure}

\begin{figure}[tbp]
\includegraphics[width=.45\textwidth]{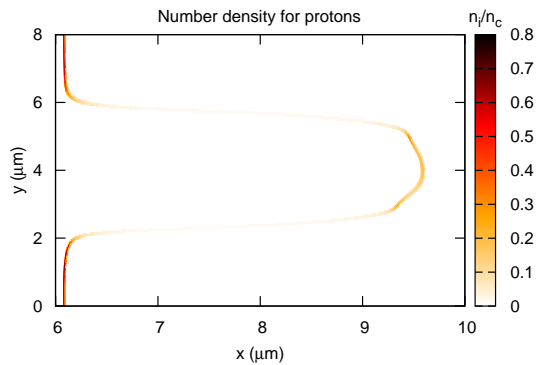}
\caption{(Color online) Density distribution of protons at instant $21T_{2}$%
. }
\end{figure}

\begin{figure}[tbp]
\includegraphics[width=.43\textwidth]{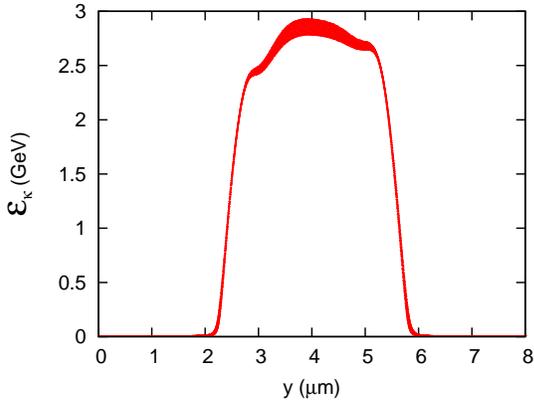}
\caption{(Color online) The final kinetic energy distribution of carbon ions
versus the transversal position $y$.}
\end{figure}

\begin{figure}[tbp]
\includegraphics[width=.43\textwidth]{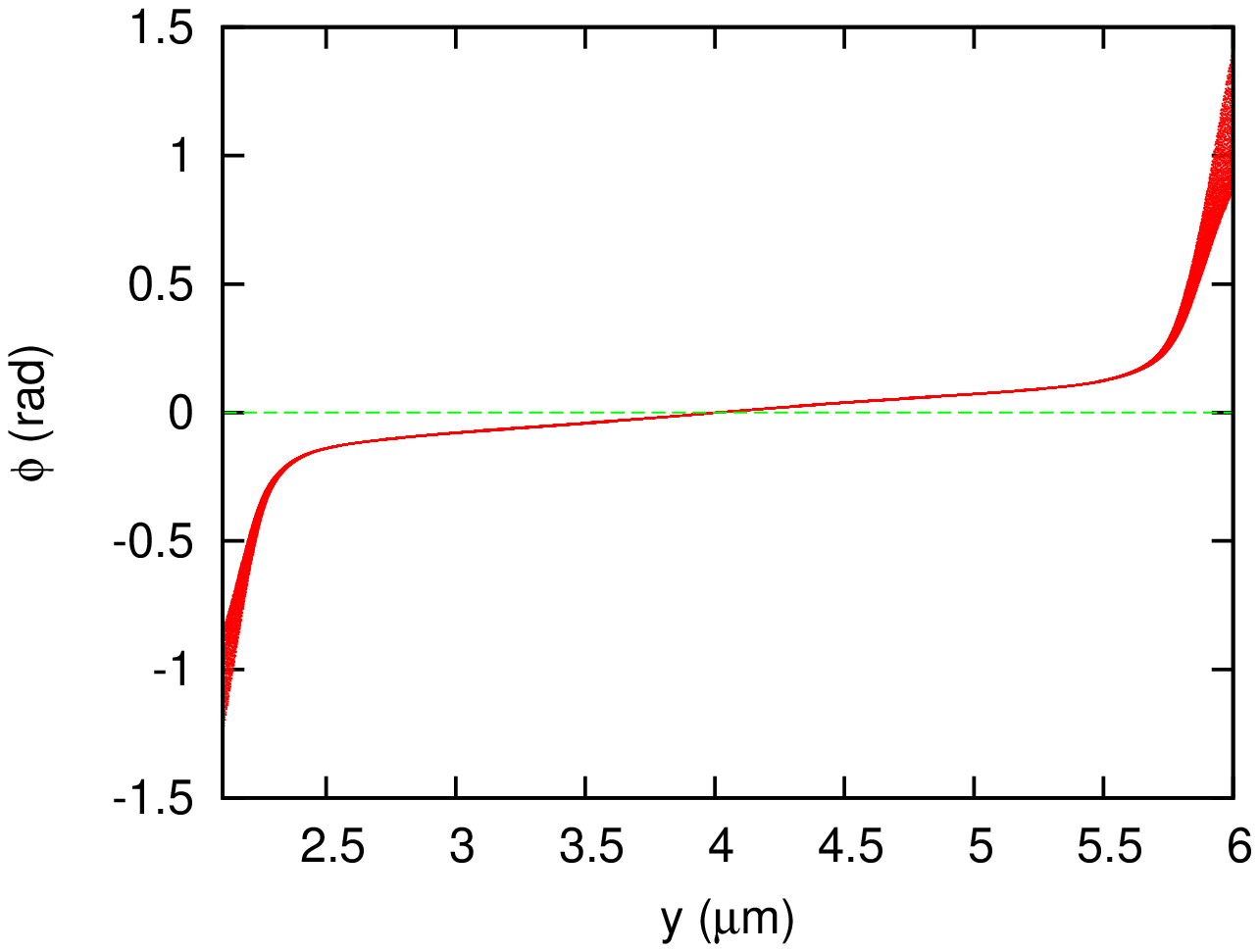}
\caption{(Color online) The angular distribution of accelerated carbon ions
versus the transversal position $y$.}
\end{figure}

\begin{figure}[tbp]
\includegraphics[width=.45\textwidth]{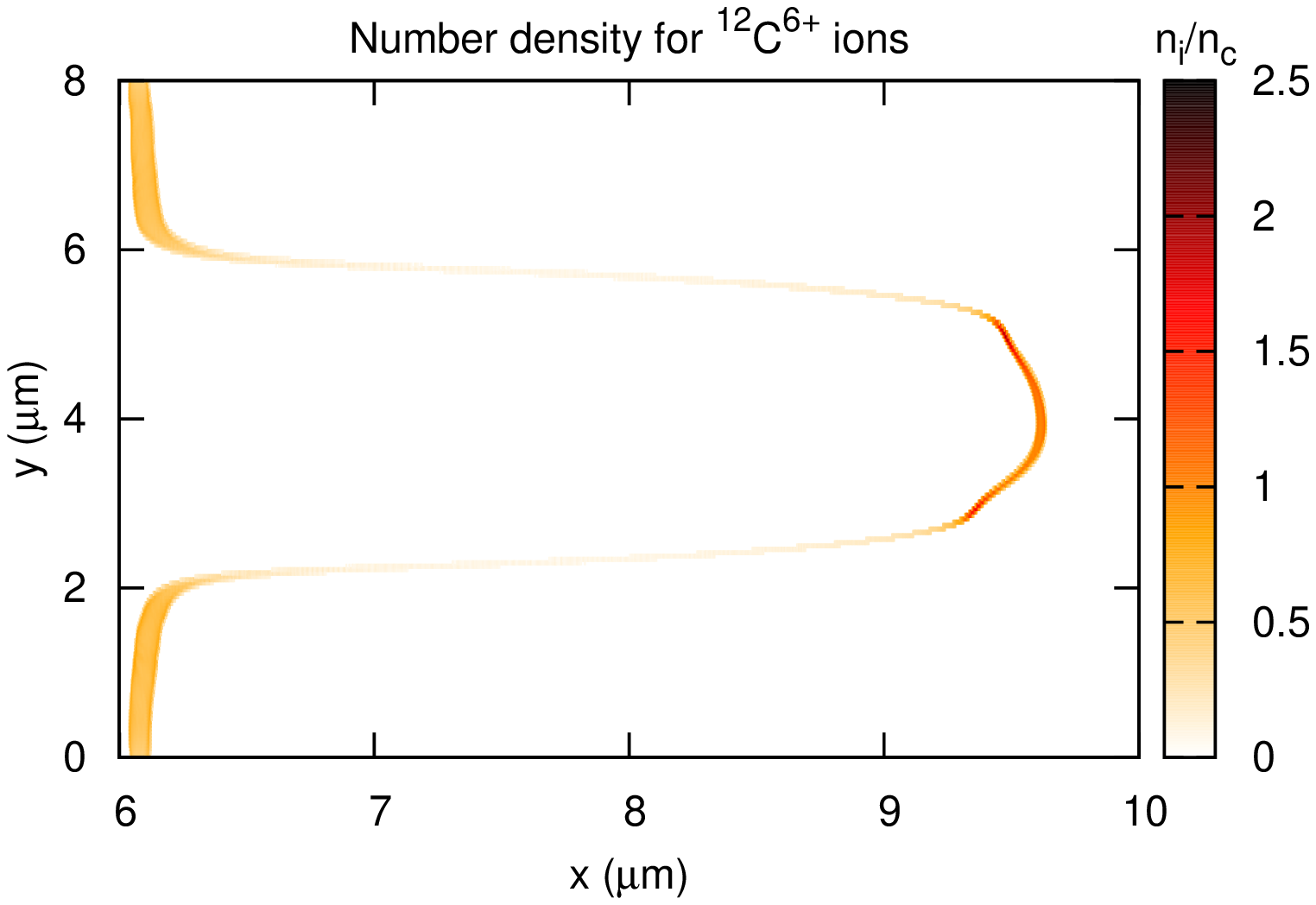}
\caption{(Color online) Density distribution for carbon ions at instant $%
21T_{2}$.}
\end{figure}

The laser pulses have profiles $\sin ^{2}\left( \pi t/\mathcal{T}%
_{1,2}\right) $ with pulse durations $\mathcal{T}_{1}=5\lambda _{1}/c$, $%
\mathcal{T}_{2}=3\lambda _{2}/c$ and Gaussian transverse profiles: $\exp
(-r^{2}/w_{1,2}^{2})$, with the waists $w_{1}=6\lambda _{1}$ and $%
w_{2}=3\lambda _{2}$. The carrier-envelope-phase of the lasers is set to
zero, so that the electric fields' maximums are at the pulses' centers. The
first laser ($\omega _{1}$) is introduced at the left boundary and
propagates along the $x$ axis from left to right, is focused at the target
layer. The second laser ($\omega _{2}$) is introduced at the right boundary
and is also focused at the target layer. We consider case $\lambda
_{1}=\lambda _{2}/2$ when $\lambda _{2}=800\ \mathrm{nm}$ which for the
phase velocity of the slowed wave gives: $\beta _{ph}=1/3$ and according to
Eq. ((\ref{gain})) for reflected particle normalized kinetic energy gives: $%
\gamma _{f}-1\approx 0.25$. Before PIC simulation we have cleared up the
role of initial conditions. It is evident that for the short laser pulses,
depending on the initial position of an ion, reflection will take place with
various longitudinal velocities (acquired in the fields), resulting to
different final energies. This is well seen from Fig. 1, displaying the role
of initial conditions: the final energy versus the initial position $x_{0}$
of an ion for plane waves, at $\mathbf{v}_{0}=0$. The intensity of effective
slowed wave is above the critical point: $\Xi _{1}=0.6$, $\Xi _{2}=0.3$.
Numerical results show that this model is almost accurate in predicting the
final energies of reflected particles. Thus, taking into account the result
for PIC simulations, the target layer is placed at the distance of $%
15.2\lambda _{1}$ from the left boundary, and the first laser is introduced
with time delay to start pulse: $t_{del}=10\lambda _{1}/c$.

In Fig. 2 ion layer blown out from a thin $^{1}\mathrm{H}^{+}$ target of
thickness $8\ \mathrm{nm}$\ by counter-propagating waves, is shown for
various instants. Hear $n_{e}=n_{c}$, where as a measure of density we take
critical plasma density $n_{c}=1.74\times 10^{21}\ \mathrm{cm}^{-3}$
calculated for $800\ \mathrm{nm}$ laser. Note that electrons are escaped
from the target before the arrival of the pulses' peaks. At the $21T_{2}$
the ions are already free. The final kinetic energy $\mathcal{E}_{k}$ and
angular $\phi =\tan ^{-1}\left( \Pi _{x}/\Pi _{y}\right) $ distributions of
accelerated protons versus the transversal position, are shown in Figs. 3
and 4, respectively. The number density of protons is displayed in Fig. 5.
The results for carbon foil of density $n_{e}=80n_{c}$ and $4\ \mathrm{nm}$
thickness are shown in Figs. 6, 7, 8. Estimates for transverse ($\epsilon
_{t}$) and longitudinal ($\epsilon _{l}$) emittances of the ions bunch show
that for protons $\epsilon _{t}$ $<$ $0.1\pi $ $\mathrm{mm}$ $\mathrm{mrad}$
and $\epsilon _{l}<10^{-8}\ \mathrm{eV\ s}$, while for carbon ions $\epsilon
_{t}$ $<$ $0.1\pi $ $\mathrm{mm}$ $\mathrm{mrad}$ and $\epsilon
_{l}<10^{-7}\ \mathrm{eV\ s}$. These results are on several orders of
magnitude smaller than their counterparts in conventional ion accelerators.
The energy spreads are: $\delta \mathcal{E}/\mathcal{E}\sim 10^{-2}$.

In conclusion, owing to the threshold character of particle reflection
phenomenon, efficient generation of quasimonoenergetic and low emittance
fast ion bunches of solid densities from nanotargets by counterpropagating
laser pulses of ultrarelativistic intensities has been put forward and
demonstrated with PIC simulations. Particularly, a setup is considered which
provides generation of ion bunches in the range of such energies and low
spreads that may have application in hadron therapy.

\begin{acknowledgments}
This work was supported by SCS of RA.
\end{acknowledgments}

\end{document}